\documentclass[%
 reprint,
 amsmath,amssymb,
 aps,
]{revtex4-1}

\usepackage{graphicx}
\usepackage{dcolumn}
\usepackage{bm}

\begin{document}

\preprint{APS/123-QED}

\title{Crosslinker mobility governs fracture behavior of catch-bonded networks}
\author{Jos\'e Ruiz-Franco}
\affiliation{Physical Chemistry and Soft Matter, Wageningen University \& Research, Stippeneng 4, 6708WE Wageningen, Netherlands.}
\author{Justin Tauber}%
\affiliation{Physical Chemistry and Soft Matter, Wageningen University \& Research, Stippeneng 4, 6708WE Wageningen, Netherlands.}
\author{Jasper van der Gucht}%
\email[Corresponding author: ]{jasper.vandergucht@wur.nl}
\affiliation{Physical Chemistry and Soft Matter, Wageningen University \& Research, Stippeneng 4, 6708WE Wageningen, Netherlands.}

\date{\today}

\begin{abstract}
While most chemical bonds weaken under the action of mechanical force (called slip bond behavior), nature has developed bonds that do the opposite: their lifetime increases as force is applied. While such catch bonds have been studied quite extensively at the single molecule level and in adhesive contacts, recent work has shown that they are also abundantly present as crosslinkers in the actin cytoskeleton.  However, their role and the mechanism by which they operate in these networks have remained unclear. Here, we present computer simulations that show how polymer networks crosslinked with either slip or catch bonds respond to mechanical stress.  Our results reveal that catch bonding may be required to protect dynamic networks against fracture, in particular for mobile linkers that can diffuse freely after unbinding. While mobile slip bonds lead to networks that are very weak at high stresses, mobile catch bonds accumulate in high stress regions and thereby stabilize cracks, leading to a more ductile fracture behavior. This allows cells to combine structural adaptivity at low stresses with mechanical stability at high stresses.

\end{abstract}

\maketitle
Many natural and engineering materials need to combine mechanical stability with structural adaptivity. Such seemingly contradictory properties can be realized in transient polymer networks: networks that are connected by dynamic, reversible bonds. The short-lived character of individual bonds allows for rearrangements and plasticity, while the mechanical integrity of the whole network can be maintained by distributing  mechanical stresses over many bonds. This leads to viscoelastic behaviour, and the possibility for stresses to relax and for damage to spontaneously heal~\cite{skrzeszewska2010fracture}. Biological examples of transient networks can be found in the  cytoskekeleton, where long actin filaments are linked together by a large variety of dynamic crosslinkers~\cite{lieleg2009cytoskeletal,huber2015cytoskeletal}, and in the extracellular matrix, where protein fibrils and polysaccharides are crosslinked into a complex network by non-covalent interactions~\cite{mak2020impact}. Synthetic examples of transient networks are associative polymers that carry hydrophobic sticky groups~\cite{mayumi2016fracture}, hydrogen-bonding groups~\cite{lewis2014influence}, or ionic interactions~\cite{henderson2010ionically,lemmers2010multiresponsive}.

The mechanical stability of transient networks relies on a balance between bond breaking and reformation events. However, because forces acting on the linkers influence their binding and unbinding rates, this balance is shifted by mechanical stress.  Local unbinding events can lead to small defects or microcracks in the material, which tend to concentrate stresses. Most crosslinkers unbind faster when force is applied, so that bond rupture is enhanced near defects, destabilizing these regions even further.  This cascade of force-induced bond rupture  ultimately leads to crack initiation and fracture, thereby compromising the resistance of transient networks against mechanical stress~\cite{skrzeszewska2010fracture,van2018laser}. 

\begin{figure}[!t]
\includegraphics[width=\linewidth]{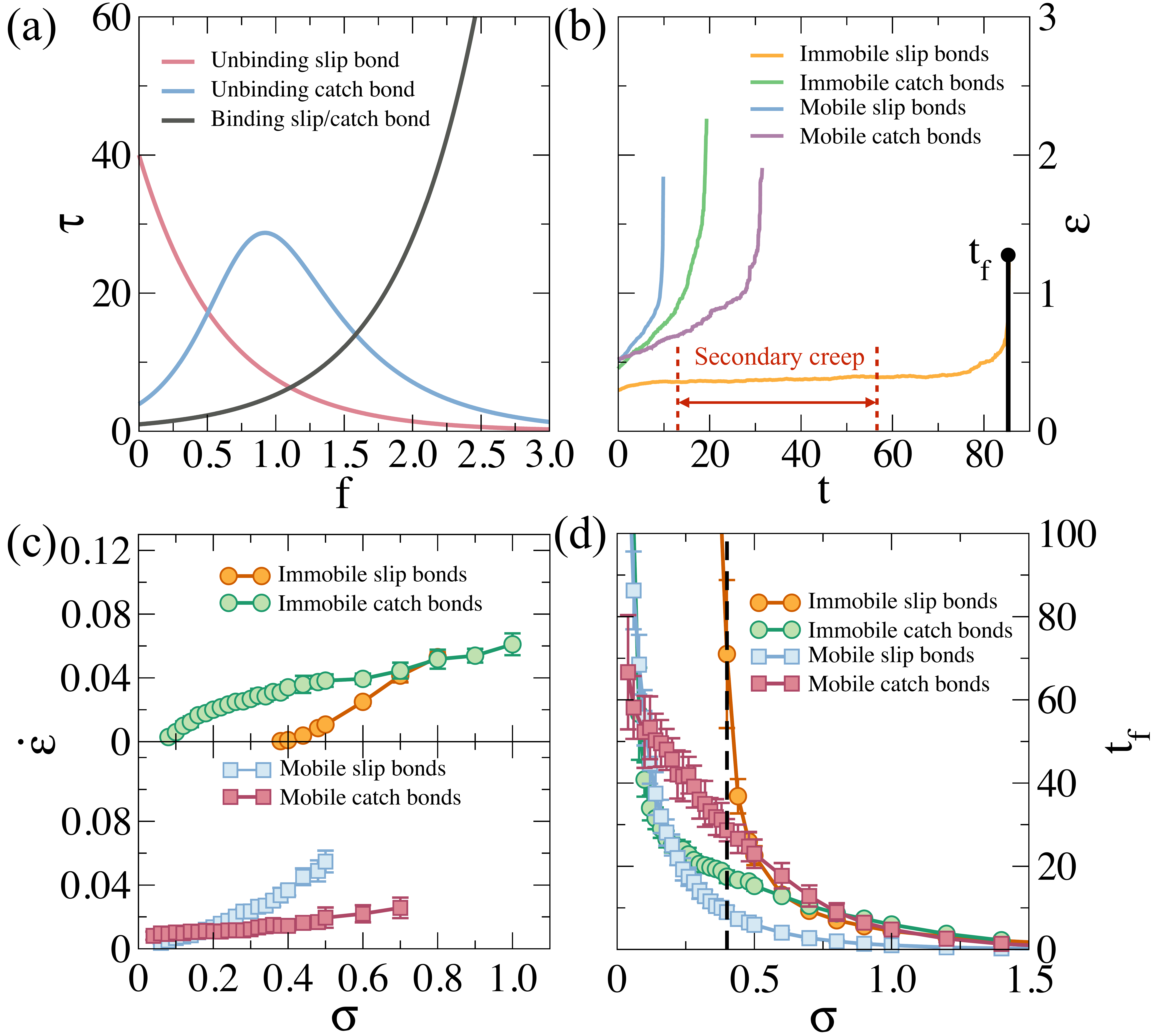}
\caption{(a) Unbinding times $\tau_{u}=1/k_{u}$ and binding times $\tau_{b}=1/k_{b}$ for slip and catch bonds. (b) Strain as a function of time for the four different types of crosslinkers, for $\sigma=0.40$. The secondary creep phase and the network lifetime $t_{f}$ are indicated for the immobile slip bond simulations. (c) Secondary creep rate and (d) lifetime of the network before fracture as a function of $\sigma$ for the four networks simulated here. The vertical dashed line indicates the onset of metastability for the immobile slip bonds, for which no fracture was observed during our simulations.}
\label{fig:Fig1}
\end{figure}

Recent findings suggest that nature may have found a way to avoid this catastrophic cascade of bond disruption, by making use of so-called catch bonds.  Instead of weakening under force (called slip bond behaviour), catch bonds first become stronger when they are stressed and weaken only at higher forces (see Fig. \ref{fig:Fig1}(a)). This counter-intuitive behavior emerges from a conformational rearrangement within the molecule upon the application of a mechanical force. First discovered in adhesion proteins~\cite{marshall2003direct,liu2020high,liu2014accumulation,luca2017notch,thomas2002bacterial}, catch bonding has recently also been demonstrated in several crosslinking proteins in the cytoskeleton~\cite{huang2017vinculin,laakso2008myosin,yao2013stress,mulla2020weak,doss2020cell}. Since catch bonds become more stable at moderate forces, they may be able to stabilize networks against fracture by accumulating in regions of high stress, and thereby mitigate the vulnerability of transient networks to defects and cracks.

As suggested by recent simulations, the mechanical stability of transient networks  depends sensitively on the mobility of the crosslinkers. Mobile linkers, such as actin-binding proteins that can diffuse freely after unbinding, can rebind in new locations of the network, and so redistribute rapidly. Immobile crosslinkers, however, such as pendant sticky groups attached to the polymer backbone, can only rebind in the same location. For slip bonds, crosslinker mobility is expected to accelerate fracture, because dissociated bonds can diffuse away from the crack tip and rebind in regions of lower stress~\cite{mulla2018crosslinker}. By contrast, mobile catch bonds may be able to stabilize networks more efficiently than immobile catch bonds, because they can accumulate more efficiently in regions of high stress~\cite{mulla2020weak}.

To test this hypothesized stabilization mechanism of catch bonds, and to investigate how the binding and unbinding kinetics of transient slip and catch bonds and their mobility couple to the distribution of stress in the network, we perform computer simulations. Biological polymer networks are highly disordered and  heterogeneous, which leads to very heterogeneous stress distributions and strongly non-affine deformation fields~\cite{arevalo2015stress,liang2016heterogeneous,shivers2019normal,ruiz2022force,zhang2017fiber,dussi2020athermal}. We therefore use a network model in which the connectivity and topology are explicitly accounted for and in which non-affine deformations and inhomogeneous stresses arise naturally. We consider a 2D network consisting of $L\times L$ nodes that can be connected by linear Hookean springs with force-extension relation $f=\mu(l-l_0)/l_0$. We set the stiffness $\mu$ and rest length $l_0$ to unity for all springs, so that all forces are expressed in units of $\mu$ and all length scales in units of $l_0$. The transient nature of the network is accounted for by allowing bonds between nodes to bind and unbind stochastically, using a kinetic Monte Carlo scheme~\cite{gillespie1976general}. Slip bond behavior is described using Bell's model, which assumes an exponential increase of the unbinding rate $k_u$ with force~\cite{bell1978models}, while catch bond behavior is described using the two-pathway model~\cite{pereverzev2005two}. As natural catch bonds tend to be weaker than slip bonds~\cite{mulla2020weak}, we choose parameters for which the lifetime of catch bonds at rest is much shorter than for slip bonds, as shown in Fig. \ref{fig:Fig1}(a), which shows the bond lifetimes $\tau_u=1/k_u$ as a function of force for both types. We furthermore assume that the rebinding rate $k_b$ decreases with increasing distance between the nodes, and take this to be the same for both types of bonds, as also shown in Fig. \ref{fig:Fig1}(a). All times are expressed in units of the binding time at zero force, $1/k_0^b$. To study how linker diffusion affects the mechanics of the network, we compare mobile and immobile bonds. While the binding and unbinding of immobile bonds always involve the same pair of nodes, mobile bonds are allowed to rebind at any location in the network, which corresponds to the limit of rapid diffusion after unbinding. To enable mobile bonds to redistribute, we allow new bonds to form between nodes that are already connected, i.e. we allow the formation of double bonds. The number of actually bound linkers  $N_b(t)$ can never exceed the total number of linkers $N$, which we take to be equal to the number of bonds in a network that is fully connected by single bonds. To study network failure under mechanical loading, we subject the networks to uniaxial deformation by applying a constant macroscopic stress $\sigma_{yy}\equiv\sigma$ at the top and bottom boundaries, while we use periodic boundary conditions in the $x$-direction. We assume that the network is athermal, and use the FIRE algorithm~\cite{bitzek2006structural} to minimize the potential energy after every binding or unbinding event, by adjusting the positions of the nodes in the network. Thus, mechanical stress relaxation in the networks is significantly faster than the binding/rebinding events. While our model ignores the contribution of thermal fluctuations to the mechanical response, it has been shown previously that 2D athermal network models give a very good description of biological networks of semiflexible fibrils~\cite{broedersz2016modeling}. In all cases, we start with a fully (single-)connected triangular network and perform an equilibration process during $10^{5}$ Monte Carlo steps with $\sigma=0$, before starting the deformation. A detailed description of the model is presented in the Supplemental Material. All quantities reported are averaged over $25$ independent configurations for each stress value.
 
Throughout this paper, we will compare networks with four different types of crosslinkers: immobile and mobile slip bonds, and immobile and mobile catch bonds. In Fig.~\ref{fig:Fig1}(b) we show how the four different networks deform after applying stress ($\sigma=0.40$), by plotting the macroscopic strain as a function of time. In all cases, we observe an initial elastic response, followed by a stage in which the network gradually deforms at more or less constant strain rate (denoted secondary creep), until it finally fractures into two disconnected pieces.  We compute the creep rate $\dot{\varepsilon}$ in the secondary creep stage and the time to fracture $\tau_f$, and plot these as a function of $\sigma$ in Fig.~\ref{fig:Fig1}(c) and (d). We first discuss the case of immobile bonds. Clearly, the creep rate is much larger for immobile catch bonds than for immobile slip bonds. This is a consequence of the larger intrinsic unbinding rate $k_0^u$ of catch bonds compared to slip bonds, making catch-bonded networks more dynamic. While the creep rate increases almost linearly with increasing stress for slip bonds, it levels off at intermediate stresses for catch bonds. 
 We can further interpret these data by calculating the effective creep viscosity as $\eta=\sigma/\dot{\varepsilon}$. As shown in Fig. S1, the force-enhanced unbinding kinetics for slip bonds gives rise to strain thinning behavior (i.e. a viscosity that decreases with increasing stress), while for catch bonds, the strain thinning behavior at small forces is followed by strengthening of bound linkers at larger forces, giving rise to very pronounced strain thickening (a viscosity that increases with increasing stress). When considering the network lifetime $\tau_f$, we find that for all stresses, immobile slip bonds are more efficient in postponing fracture than immobile catch bonds. In particular, for slip bonds we find a metastable region for $\sigma\lesssim0.40$ (highlighted by the dashed line in Fig.~\ref{fig:Fig1}(d)), for which no fracture is observed during our simulations and the creep rate vanishes. This metastable shifts to much lower stresses for catch bonds. These findings show that, although immobile catch bonds can qualitatively change the response of the network to stress, they are not very efficient in protecting networks against fracture: even though the lifetime of individual catch bonds is significantly larger than that of slip bonds at moderate forces (Fig. \ref{fig:Fig1}(a)), this does not lead to an increase in network lifetime at higher stresses. Furthermore, we note that the characteristic peak at intermediate forces in the lifetime of individual catch bonds is not visible in the lifetime of the networks, indicating that collective effects effectively screen the catch bond behaviour. As we show in Fig. S2, this screening is related to the transient nature of the bonds, as a peak in network lifetime does appear in the absence of rebinding (i.e. for $k_{b}=0$).
 
 The picture changes dramatically, however, when the bonds are mobile, so that they can redistribute within the network. As shown in Fig. \ref{fig:Fig1}(c), crosslinker mobility strongly enhances the creep rate for slip-bonded networks. This enhanced creep significantly accelerates fracture, and removes the metastable regime that was observed for immobile slip bonds (Fig. \ref{fig:Fig1}(d)). Such an adverse effect of mobility on the stability of slip bonded materials was also observed in 1D simulations of adhesive patches,~\cite{mulla2018crosslinker}. Interestingly, for catch bonds mobility has the opposite effect: it  reduces the rate of creep and delays fracture. Catch bonding thus provides a mechanism to stabilize networks with dynamic, diffusable linkers, such as those in the cytoskeleton. In particular, mobile catch bonds can combine deformability and adaptivity at low stresses (note that for $\sigma\lesssim0.10$ the creep rate significantly exceeds that of slip bond networks), with rigidification and stabilization at higher stresses.

\begin{figure}[!t]
\includegraphics[width=\linewidth]{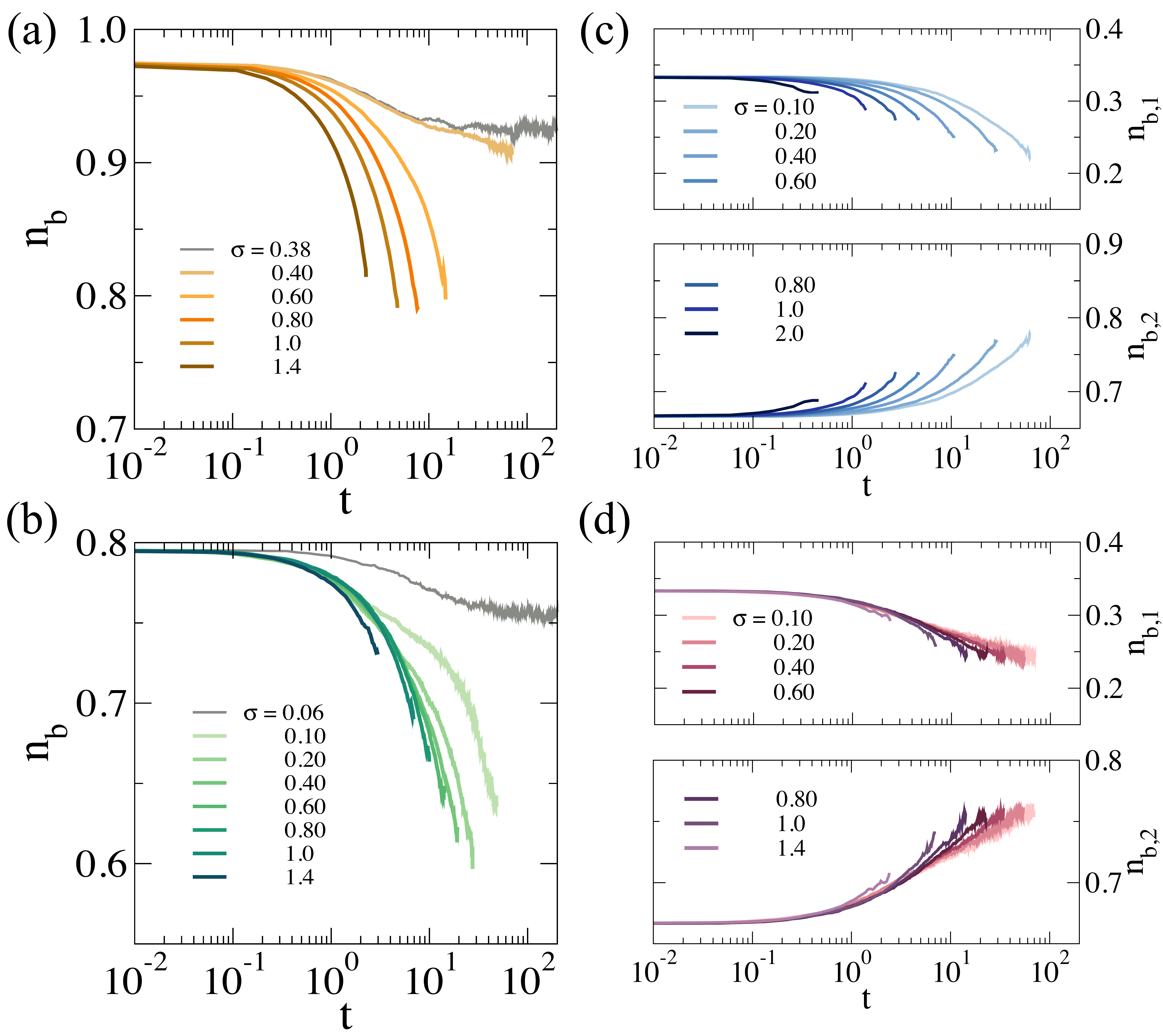}
\caption{Evolution of the normalized number of bonds in the network for different stresses $\sigma$, shown as $n_{b}=N_{b}(t)/N$ with $N_b(t)$ the number of bonds present at time $t$ and $N$ the maximum number of bonds, for (a) immobile slip bonds, (b) immobile catch bonds, (c) mobile slip bonds, and (d) mobile catch bonds. For the mobile bonds, the evolution of single bonds $n_{b,1}$ and double bonds $n_{b,2}$ is shown separately. The end of each curve (except for the metastable case in (a) and (b)) denotes the moment of fracture.}
\label{fig:Nb}
\end{figure}

To obtain  microscopic insight in the underlying crosslinker dynamics, we consider the evolution of the number of bonds in the network $n_b$ as a function of time. For immobile slip bonds, shown in Fig.~\ref{fig:Nb}(a), $n_{b}$ decays gradually after applying stress, and, as expected based on the lifetime-force relation of the bonds, this decay is faster when the stress increases. The accumulation of ruptured bonds gradually weakens the network, which makes it more stretchable and causes the secondary creep observed  in Fig. \ref{fig:Fig1}(b). The metastable state for low stresses corresponds to a plateau in $n_b$ (grey curve in \ref{fig:Nb}(a)), in which bond rupture is balanced by rebinding, so that small cracks can be repaired before they start to propagate. For immobile catch bonds, we see a similar behaviour as for slip bonds, although the decay rate is less dependent on the stress Fig.~\ref{fig:Nb}(b)). While this shows that the catch mechanism is indeed activated in these networks, it does not work very efficiently, because the lifetime of the network remains smaller than that of slip-bonded networks (Fig. \ref{fig:Fig1}(d)).  The scenario changes for the case of mobile bonds. As shown in the upper panels of Fig. \ref{fig:Nb}(c) and (d), the number of single bonds decreases similarly as for immobile bonds. However, this decrease is now accompanied by an increase in the number of double bonds (shown in the lower panels). This increase of $n_{b,2}$ indicates that the dissociated bonds   rebind in other regions of the network. The total number of bonds, $n_{b,1}+n_{b,2}$, remains more or less constant (Fig. S3), but their distribution over the network is strongly affected by the applied stress. For slip bonds, this bond redistribution destabilizes the network, in particular for higher stresses, since the network ruptures already after a small change in $n_{b,1}$ and $n_{b,2}$. For catch bonds, however, this is not so, indicating that the new bonds are placed in regions where they stabilize the network, i.e. in regions where the stress is higher. This can also be seen by looking at the probability distribution of the forces at which the bonds break and reform, shown in Fig. S5. While the  distribution for rebinding is very similar for the two different bonds, slip bonds that are more stretched break much more rapidly than stretched catch bonds.

\begin{figure*}[!t]
\includegraphics[width=\linewidth]{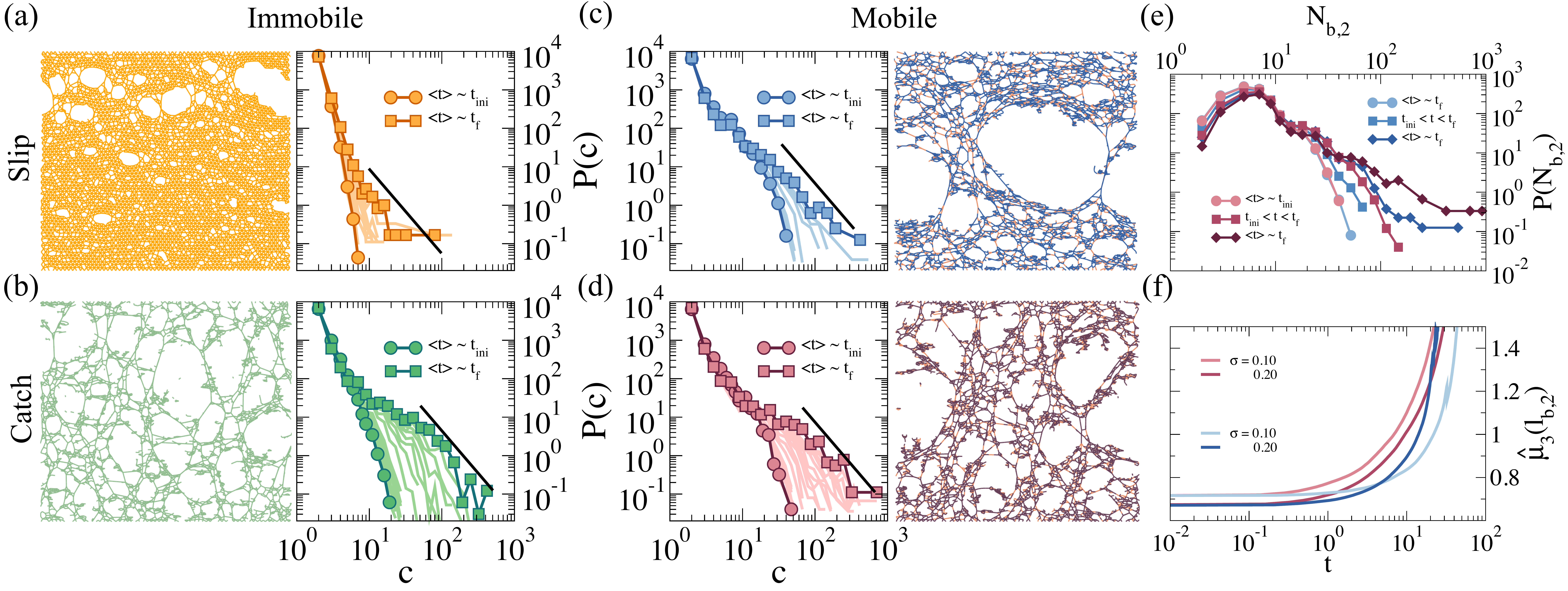}
\caption{Snapshots of the networks just before macroscopic fracture, and the evolution of the crack size distribution $P\left(c\right)$ at stress $\sigma=0.40$ for (a) immobile slip bonds, (b) immobile catch bonds, (c) mobile slip bonds, and (d) mobile catch bonds. For networks with mobile bonds, orange colour corresponds to single bonds, while blue and purple indicate double slip and catch bonds, respectively. In each panel, a slope of $-2.2$  is indicated. The initial and final $P\left(c\right)$ are indicated by circles and squares, respectively, and the intermediate times are shown as drawn lines. (e) Distribution of double bonds located at the crack surface, and (f) skewness $\hat{\mu}_{3}$ of the double bond length distribution $l_{b,2}$ for networks with slip bonds (blue) and catch bonds (red), for different values of $\sigma$.} 
\label{fig:Damage}
\end{figure*}

Next, we study how the binding and unbinding events affect the evolution of the structure of the network. Representative snapshots of the networks just before macroscopic fracture are shown in Fig. \ref{fig:Damage}(a)-(d) for the four different crosslinkers at $\sigma=0.40$. How these networks evolve in time, starting from an initially homogeneous network, is shown in Figs. S10 and S11. In all networks, we see the formation of cracks that open up due to the applied tensile force. We calculate the distribution of crack sizes $P(c)$ from these snapshots, as explained in detail in the Supplemental Information. The results are shown in Fig.~\ref{fig:Damage} next to the snapshots. Both the snapshots and the crack size distributions reveal different fracture scenarios for slip and catch bonds. For networks with slip bonds, the crack size distribution remains narrow, and the damage is localized in a few large cracks that will eventually merge, leading to macroscopic failure (Fig.~\ref{fig:Damage}(a) and (c)). For catch bonds, the crack size distribution is much wider and it develops a power law tail with an exponent $\alpha= -2.2$, which is close to the value expected for random percolation~\cite{sciortino2005one}. These findings suggest that the fracture process proceeds in a different manner for the different types of bonds. For slip bonds, stress concentration destabilizes especially the larger cracks, because the stress is largest at the tip of the largest cracks~\cite{Inglis1913stresses} and, as a consequence, the rupture rate is highest there. This leads to a scenario of crack nucleation and propagation. By contrast, catch bonds stabilize these high stress regions, and thereby prevent crack propagation. This leads to a broader crack size distribution and a fracture scenario that looks more like damage percolation~\cite{Shekhawat2013from}.
As our previous findings indicate, this stabilizing effect is much more effective for mobile bonds, because mobility allows these bonds to accumulate near the crack tips. To verify this, we calculate the distribution of double bonds located on the surface of the cracks, $P(N_{b,2})$. As shown in Fig. \ref{fig:Damage}(e), this distribution is initially similar for slip and catch bonds, but as the damage process proceeds and stresses increase, the distribution develops a much longer tail for the catch bonds. This shows that the increase in the  number of double bonds observed in Fig. \ref{fig:Nb}(d) is caused by an accumulation of catch bonds at the crack surface, where the local stress is highest. This accumulation of catch bonds in high stress-regions is further confirmed by looking at the skewness of the bond length distribution $\hat{\mu}_{3}$. A larger positive skewness indicates a more asymmetric distribution with a longer tail of strongly stretched bonds. As shown in Fig. \ref{fig:Damage}(f), $\hat{\mu}_{3}$ indeed increases much more steeply for catch bonds than for slip bonds, again indicating their accumulation in high stress regions. In Figs. S8 and S9, we show that similar results are obtained at smaller stress levels.

\begin{figure}[t]
\includegraphics[width=\linewidth]{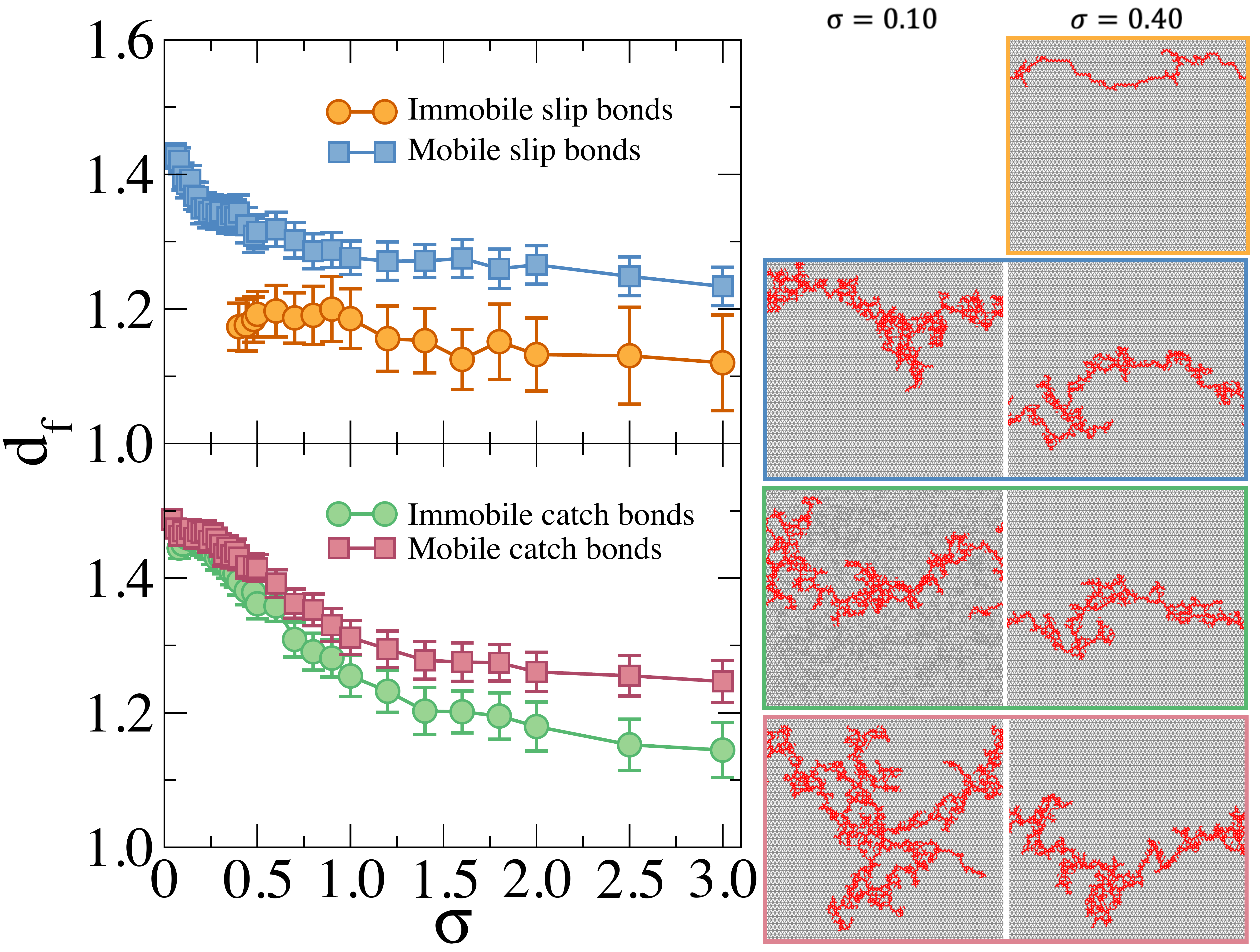}
\caption{(a) Fractal dimension $d_{f}$ of the percolating crack at $t_{f}$, mapped on the undeformed network. (b) Snapshots showing the corresponding fracture patterns on the undeformed network for (from top to bottom) immobile slip bonds, immobile catch bonds, mobile slip bonds and mobile catch bonds at $\sigma=0.10$ and $0.40$. Broken bonds belonging to the percolated crack are highlighted in red color. Note that the network with immobile slip bonds does not fracture for $\sigma=0.10$. }
\label{fig:df}
\end{figure}

As discussed above, the fracture process appears to proceed differently for the different types of bonds. To investigate this in more detail, we calculate the fractal dimension $d_{f}$ of the final percolating crack, using the box-counting method~\cite{gagnepain1986fractal,ruiz2020tuning}. In Fig.~\ref{fig:df}(a) we show how $d_{f}$ varies with the stress for the different bonds. Snapshots of the final fracture patterns for two different stresses are shown in Fig.~\ref{fig:df}(b), with the percolating crack shown in red.  For immobile slip bonds, $d_{f}$ depends only weakly on the stress and has a values close to 1, which is the value expected for linear crack propagation. This can be seen also in the snapshot in Fig.~\ref{fig:df}(b), and corresponds to brittle fracture behaviour~\cite{moreira2012fracturing,dussi2020athermal}. We note that for small stresses there appears to be a weak maximum in the fractal dimension. Such a maximum is also seen in the total number of broken bonds at the moment of fracture (Fig. S4). We speculate that this maximum is related to the vicinity of the metastable regime and due to a transition in the fracture mechanism, from a subcritical regime at small stresses, where fracture occurs only after nucleation  of a sufficiently large defect to a rapid crack propagation regime at high stresses. For both immobile and mobile catch bonds, the fractal dimension is much larger than for the corresponding slip bonds, in particular at low stresses. This is in accordance with a scenario of damage percolation, and can also be seen by considering the total number of broken bonds at the moment of fracture, which is larger for catch bonds than for slip bonds (Figs. S4 and S12). As the stress increases, the fractal dimension decreases for the catch bond networks and becomes similar to that of slip bond networks, while the number of broken bonds decreases. This indicates a more brittle fracture at higher stress and can be explained by the fact that at high forces the catch bonds return to slip behaviour (Fig. \ref{fig:Fig1}(a)). 

In conclusion, our results highlight how crosslinker dynamics affect the stability and fracture of transient networks. Many biological materials, such as the actin cytoskeleton, are crosslinked by dynamic and mobile linkers, probably because this allows for fast dynamic control of the network properties. Our simulations show that for slip bonds, linker mobility seriously decreases the resistance of the networks to stress, while catch bonded networks become more stable when the bonds are mobile. This  suggests that mobile biological catch bonds have evolved as a strategy to realize flexibility and dynamics without compromising the resistance to mechanical stress. Most probably, cells use a combination of slip and catch bonds to tailor the mechanical response and dynamics of the cytoskeleton at various stress levels. We hope that our results will help to shed light on the complex dynamic behavior of biological networks, and that it can also inspire the design of novel synthetic materials with  mechanical properties that cannot be realized with simple transient bonds.

\begin{acknowledgments}
This work is part of the SOFTBREAK project funded by the European Research Council (Consolidator grant Softbreak, grant agreement 682782). The authors acknowledge Simone Dussi, Frans Leermakers and Martijn van Galen for  useful discussions.
\end{acknowledgments}


\clearpage

\onecolumngrid

\section*{Supplemental Material for \\ ``Crosslinker mobility governs fracture behavior of catch-bonded networks''}
\setcounter{equation}{0}
\setcounter{figure}{0}
\setcounter{table}{0}
\setcounter{section}{0}

\renewcommand\thefigure{S\arabic{figure}} 
\noindent

\section{Simulation Methods}
\subsection{Network model}
We perform simulations of 2D triangular networks of $L\times L$ nodes interconnected by one-dimensional Hookean springs with stretching modulus $\mu$ and rest length $l_{0}$. Thus, the Hamiltonian $\mathcal{H}$ of the system is

\begin{equation}
\mathcal{H}=\frac{\mu}{2l_{0}}\sum_{\left\langle ij\right\rangle}\left(l_{ij}-l_{0}\right)^{2}
\end{equation}

\noindent where the sum is taken over all bonds $\left\langle ij\right\rangle$ with $l_{ij}$ the bond length. All bonds are harmonic springs with unit stiffness and unit rest length (so $\mu=1$ and $l_0=1$). We use periodic boundary conditions in the $x$-direction, while nodes on the top and bottom boundaries are fixed. After each deformation step or binding/unbinding event, the energy of the system is minimized by adjusting the positions of the nodes, using the FIRE algorithm~\cite{bitzek2006structural}. The maximum tolerance $F_{RMS}$ corresponds to the system root-mean-square force and it is set to be $10^{-5}$. Since we only consider potential energy, our system is under athermal conditions. Node pairs that have no bond connected do not interact at all. Uniaxial deformation of the network in the $y$-direction leads to a macroscopic  stress in the network $\sigma$. The stress is computed by the virial stress tensor defined as

\begin{equation}
    \sigma_{\alpha\beta}=\frac{1}{A}\sum_{\left\langle ij \right\rangle}f_{ij,\alpha}r_{ij,\beta}\,,
\end{equation}

\noindent where $\alpha$ and $\beta$ are the Cartesian axes. The  sum runs over all the bonded pairs of nodes $\left\langle ij\right\rangle$, $f_{ij,\alpha}$ is the force acting on the node $i$ due to $j$ in the $\alpha-$direction, $r_{ij,\beta}$ is the distance between the two nodes in the $\beta-$direction, and $A$ is the instantaneous area. We evaluate the $\sigma_{yy}$ tensor component. In our simulations,  we maintain a constant uniaxial stress $\sigma=\sigma_{yy}$ by adjusting the strain.

All quantities are expressed in reduced units, i.e. length scales are expressed in terms of $l_0$, forces in terms of $\mu$, and stresses in terms of $\mu/l_0^2$. Time is expressed in units of the rebinding time $1/k_0^b$ (which is the same for slip and catch bonds, see below). 

\subsection{Unbinding and binding bonds}

Structural changes in the network due to binding and unbinding events are simulated by using a kinetic Monte Carlo scheme due to Gillespie~\cite{gillespie1976general}. At every step, we make a list of all possible unbinding and binding events. Based on the list we calculate the total reaction propensity $K_{tot}=\sum_{i}k_{i}$, where $k_{i}$ is the transition rate of a spring corresponding to an unbinding or binding event, as defined below. Which event is the next reaction to occur is drawn randomly, witch each event weighed with its actual reaction rate. The time interval to the next event is drawn also randomly, assuming exponentially distributed events, i.e. $\Delta\tilde{t}=-\textrm{ln}\left(\textrm{rand}\left(0,1\right) \right)/K_{tot}$, with the dimensionless time then given by $t=\tilde{t} k_{0}^{b}$.

All bonds in our networks are transient, reversible bonds with kinetics of binding and unbinding  that depend on the force acting on the bond, $f = \mu\left(l_{ij}-l_{0}\right)/l_{0}$. For the slip bonds, we assume an unbinding rate that decreases exponentially with the force, as predicted by the Bell model~\cite{bell1978models} :

\begin{equation}
k_{u}^{s}\left(f\right)=k_{0}^{s}\exp\left[\frac{f}{f_{e}^{s}}\right]\,,
\end{equation}

\noindent where $k_{0}^{s}$ is the unforced unbinding rate, and $f_{e}^{s}$ is the force where the off rate has fallen with a factor $1/e$. This is related to the activation length $\delta$ of the bond, as $f_{e}^{s}=k_BT/\delta^s$. 
By contrast,  catch bond behavior is described by the two path-way model as~\cite{pereverzev2005two}

\begin{equation}
k_{u}^{c}\left(f\right)=k_{0,1}^{s}\exp\left[\frac{f}{f_{e,1}^{s}}\right]+k_{0,2}^{c}\exp\left[-\frac{f}{f_{e,2}^{c}}\right]\,.
\end{equation}

\noindent The first term, characterized by $k_{0,1}^{s}$ and $f_{e,1}^{s}$, captures the dissociation along a slip-like path at large pulling force $f$, while the second term, characterized by $k_{0,2}^{c}$ and $f_{e,2}^{c}$ describes dissociation along a catch-path decreasing the unbinding rate (and thus, increasing the lifetime) with force for small forces $f$. To mimic catch-bond behavior, it is necessary that $k_{0,1}^{c} > k_{0,2}^{c}$ and $f_{e,1}^{c} < f_{e,2}^{c}$~\cite{pereverzev2005two,van2020chemical}. In Fig. 1(a) in the main text, we show the bond lifetime $t_{u}=1/k_{u}$ as a function of $f$ for both slip and catch bonds. In particular, we decide to fix $\left[k_{0}^{s}, f_{e}^{s}\right] = \left[0.025, 0.6\right]$ for slip bonds, and $\left[k_{0,1}^{s}, f_{e,1}^{s}\right] = \left[0.005,0.6\right]$ and $\left[k_{0,2}^{c}, f_{e,2}^{c}\right] = \left[0.25,0.3\right]$ for catch bonds. Finally, we consider that the binding rate is also load-dependent and the same for both types of bonds, being expressed as:

\begin{equation}
k_{b}\left(f\right)=k_{0}^{b}exp\left[-\frac{f}{f_{e}^{b}}\right]\,.
\end{equation}

\noindent Thus, we assume that the binding rate between a pair $ij$ becomes smaller with increasing distance between the two nodes. Fig. 1(a) in the main text shows the binding lifetime $t_{b}=1/k_{b}$ for the set of values $\left[k_{0}^{b}, f_{e}^{b}\right] = \left[1,0.6\right]$.

\subsection{Immobile and mobile bonds}

Immobile bonds rebind between the same pair of nodes from which they are unbound, providing memory to the system of its initial structure. By contrast, mobile bonds can form in random new locations, involving thus any two nodes. To allow bonds to redistribute, we allow a new bond to form between a pair of nodes that is already linked by one bond, so that double bonds can form.  This implies that, while for immobile bonds the maximum local connectivity is $z^{\left(i \right)}_{max}=6$, for mobile bonds  this is $z^{\left(m \right)}_{max}=2z^{\left(i \right)}_{max}$. In all our simulations, we impose a maximum number of bonds that can be formed $N$, which is  equal to the number of bonds that is present in a network where every pair of neighbouring bonds is connected with exactly one bond. This means that new bonds can only form if there are unbound linkers available.

\newpage

\section{Additional results}

\subsection{Effective viscosity}

\begin{figure*}[!h]
\includegraphics[width=0.7\linewidth]{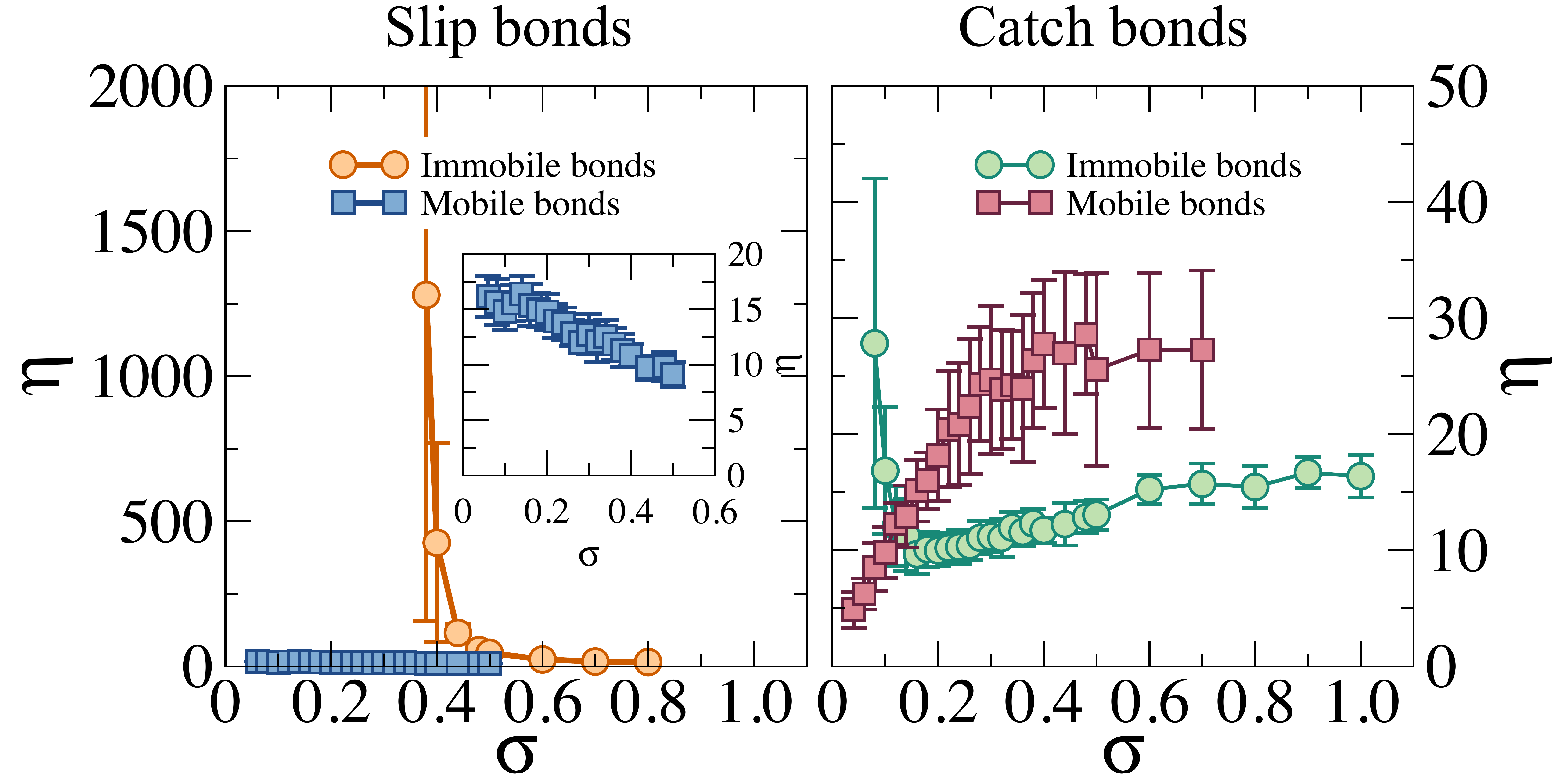}
\caption{Effective viscosity $\eta=\sigma/\dot{\epsilon}$, for networks with (a) slip bonds and (b) catch bonds as a function of the stress $\sigma$ and the mobility. \textit{Inset}: Zoom on the viscosity for mobile slip bonds.}
\label{fig:Viscosity}
\end{figure*}

Fig.~\ref{fig:Viscosity} shows the effective elongational viscosity computed as $\eta=\sigma/\dot{\epsilon}$ for networks with slip and catch bonds. For networks with slip bonds, Fig.~\ref{fig:Viscosity}(a), we observe that $\eta$
decreases with increasing stress, indicating strain thinning. For immobile bonds $\eta$ is larger than for mobile bonds, which is in line with our finding that the network lifetime is larger for immobile bonds. In particular, for immobile bonds we observe that the viscosity diverges for low stress, which corresponds to the metastable regime.

By contrast, networks with catch bonds manifest strain thickening, Fig.~\ref{fig:Viscosity}(b). Furthermore, unlike what happens for slips bonds, $\eta$ is considerably larger when catch bonds are mobile. This opposite behavior between bonds as a function of mobility strongly suggests that mobility is needed to take advantage of the nature of catch bonds. 

\subsection{Catch bond behaviour on the network}

\begin{figure*}[!h]
\includegraphics[width=0.45\linewidth]{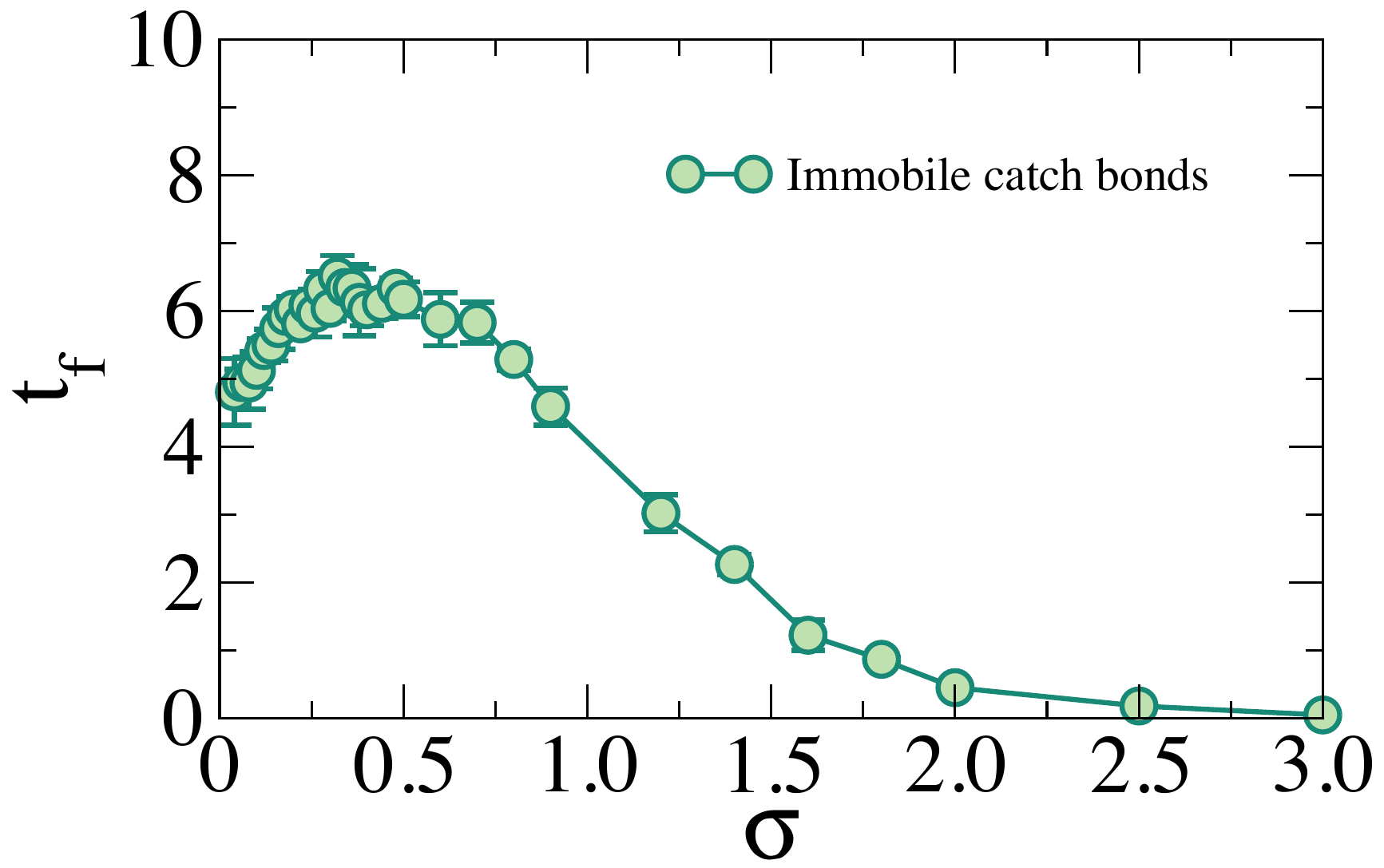}
\caption{Lifetime of a network with irreversible catch bonds  as a function of $\sigma$. Here, the rebinding rate $k_{b}$ is assumed to be zero.}
\label{fig:FigS1}
\end{figure*}

The lifetime of a network with catch bonds without rebinding transitions, i.e. $k_{b}=0$, is shown as a function of stress in Fig.~\ref{fig:FigS1}. We observe how a peak in $\tau_f$ develops at $\sigma\approx0.5$. The presence of this peak resembles the characteristic peak of an individual catch bond. However, it is less pronounced, probably because the stress distribution in the network is heterogeneous. Since  rebinding events promote bond formation at low stress, the echo from the catch bond behavior is shielded in transient networks with reversible bonds.

\subsection{Evolution of the total number of mobile bonds }

As mentioned in the main text, the total number of bonds $n_{b,1}+n_{b,2}$ for both mobile slip and catch bonds remains more or less constant, except close to the moment of fracture, when the network is destabilized and no more bonds can be bound close the cracks to postpone their propagation.

\begin{figure}[!h]
\includegraphics[width=0.9\linewidth]{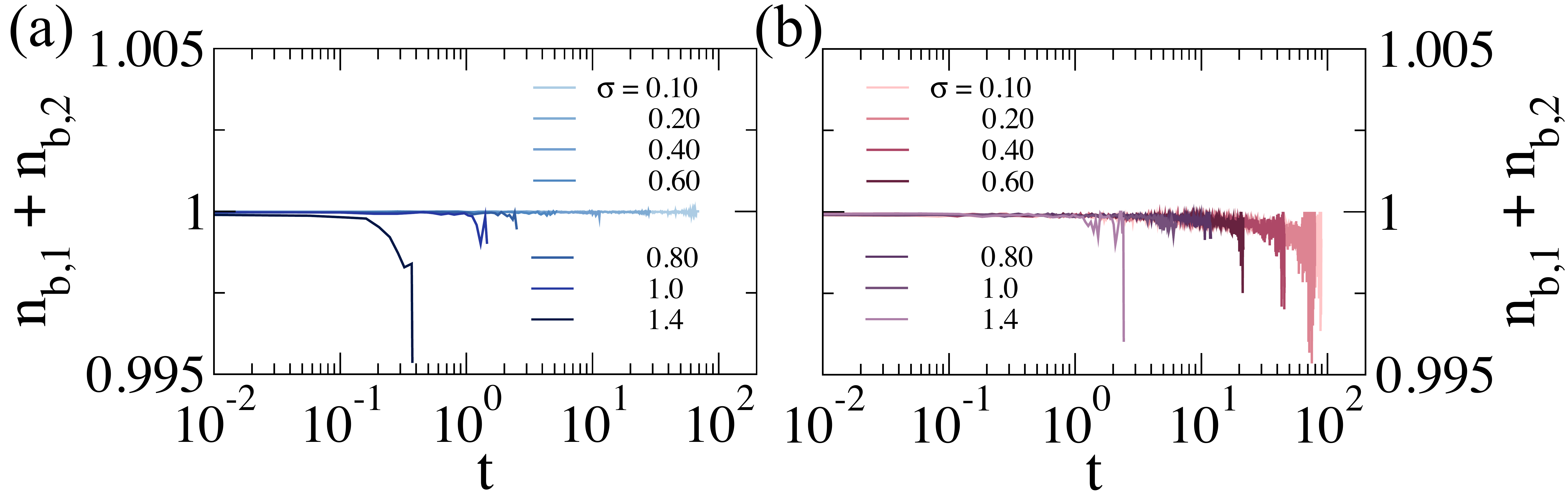}
\caption{Evolution of the total number of bonds, $n_{b,1}+n_{b,2}$ for (a) mobile slip bonds and (b) mobile catch bonds.}
\label{fig:nb1nb2}
\end{figure}

\subsection{Fraction of final bonds}
\begin{figure}[!h]
\includegraphics[width=\linewidth]{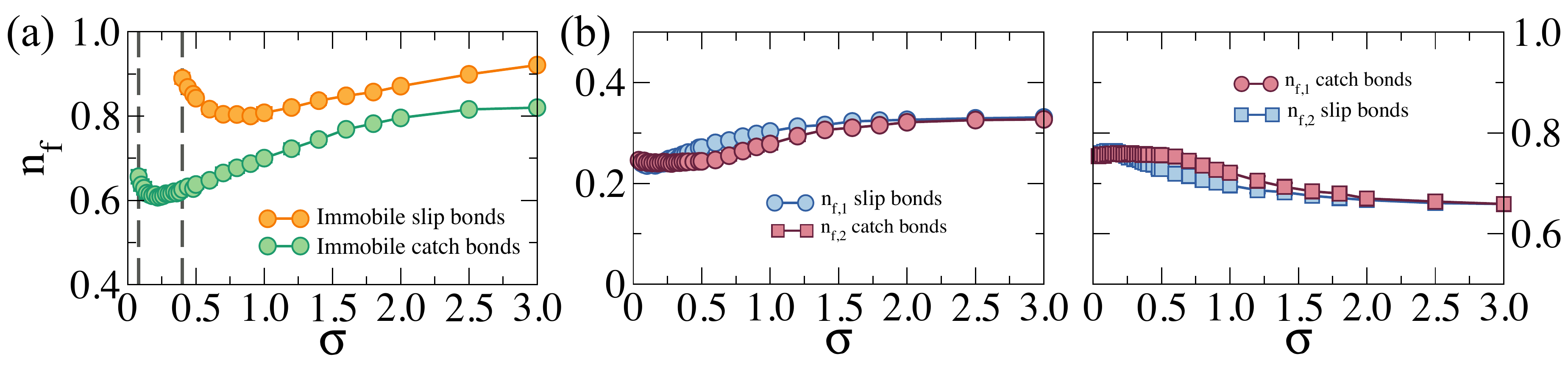}
\caption{(a) Fraction of the number of bonds at the moment of fracture, $n_{f}=N_{b}\left(t_{f}\right)/N$, for immobile slip and immobile catch bonds as a function of $\sigma$. The vertical dashed line indicates the onset of the metastability. (b) Fraction of final single $n_{f,1}$ and double $n_{f,2}$ bonds for networks with mobile slip and mobile catch bonds.}
\label{fig:Fractionfinal}
\end{figure}

In Fig.~\ref{fig:Fractionfinal}(a) we report the fraction of bonds at the moment of fracture, $n_{f}=N_{f}/N$ as a function of stress for immobile bond networks. In particular, for slip bonds, we see a minimum at $\sigma\approx0.5$, i.e. close to the metastability. The presence of this minimum hints at a change in fracture scenario, from  fracture at subcritical stress at low $\sigma$, to fracture dominated by gradual degradation of the network at high $\sigma$. This minimum is shifted to $\sigma\approx0.22$ for immobile catch bonds, leading to a  metatable state at lower stress, and $n_{f}$ is a monotonic function of $\sigma$, indicating that the fracture becomes more brittle as the stress increases. We note furthermore that the fraction of bonds remaining in the network is much smaller for catch than for slip bonds, indicating much more diffuse damage in the case of catch bonds. 

Fig.~\ref{fig:Fractionfinal}(b) shows the fraction of single $n_{f,1}$ and double $n_{f,2}$ bonds for networks with mobile slip and catch bonds. We observe that for intermediate stresses, $n_{f,2}$ is higher for catch bonds, indicating that the redistribution of bonds by the accumulation of double bonds proceeds more efficiently for catch bonds. This is caused by the fact that catch bond networks are stabilized by relocalizing bonds, while slip bond networks are destabilized,

\subsection{Unbinding and binding probability distribution}
\begin{figure*}[!h]
\includegraphics[width=0.9\linewidth]{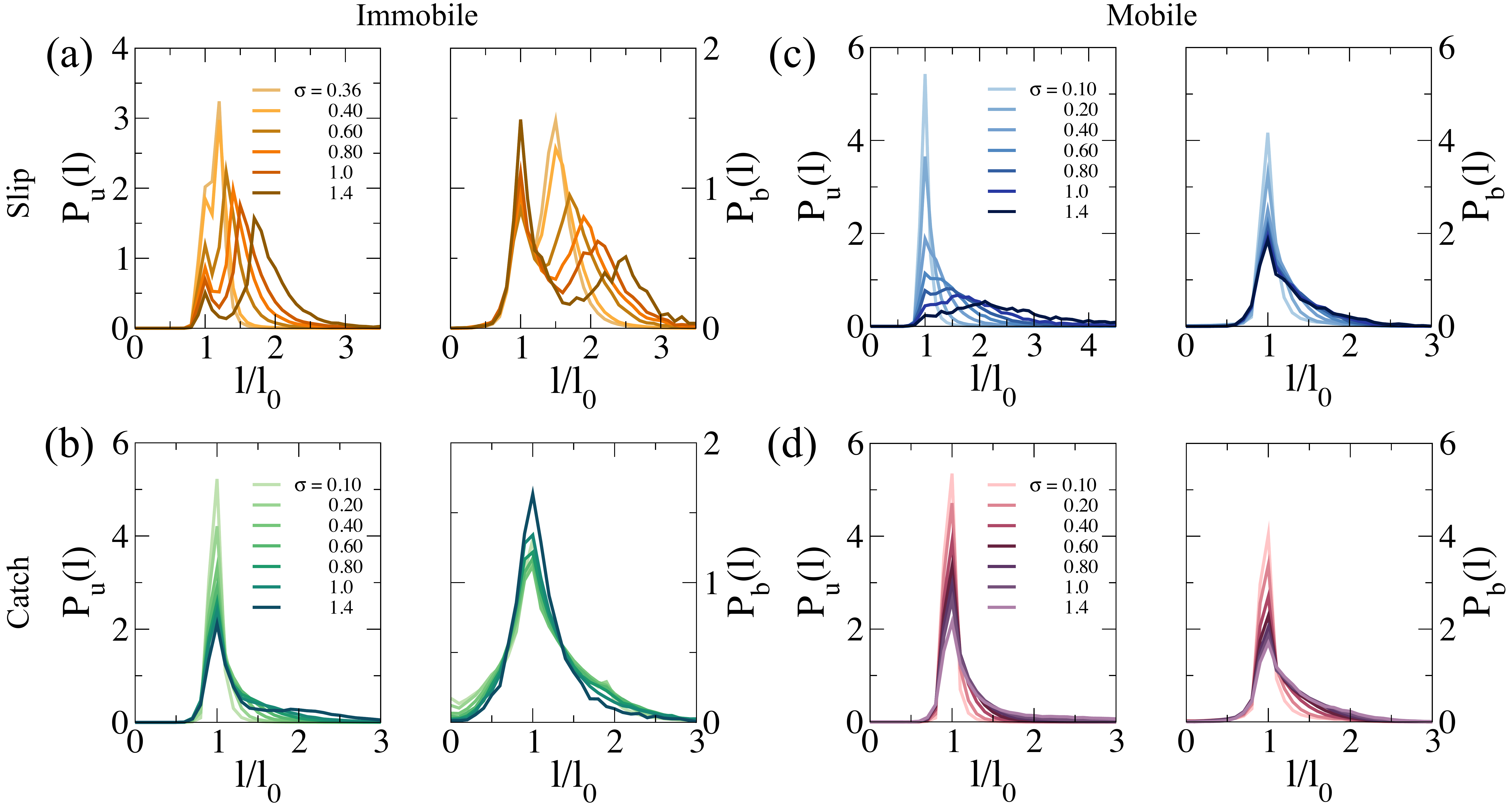}
\caption{Unbinding $P_{u}\left(l\right)$ and binding $P_{b}\left(l\right)$ probability distributions for different stress values $\sigma$. (a) Immobile slip bonds, (b) immobile catch bonds, (c) mobile slip bonds, and (d) mobile catch bonds.}
\label{fig:Prob}
\end{figure*}

We record the bond length associated to each binding/unbinding event during the simulation, and build the probability distribution that a bond will unbind $P_{u}$ or bind $P_{b}$ with length $l$. This information is shown in Fig.~\ref{fig:Prob} for the four different types of networks we have simulated here, for different $\sigma$. While the mobility of bonds strongly influences the probability distributions for slip bonds, for catch bonds they are quite similar. Indeed, we observe that both $P_{b}$ and $P_{u}$ for immobile slip bonds develop a double peak; the first placed at the bond rest length, whereas the second one shifts to higher bond length with $\sigma$. Because immobility limits the places where bonds can be formed, it is expected that slip bonds are forced to bind with $l_{ij}\gg l_{0}$. However, these strongly stretched bonds will also break rapidly due to their slip behaviour. When slip bonds are mobile, the second peak in $P_{b}$ disappears, and the distribution develops a long tail. Now, slip bonds can form wherever they want, and, since their lifetime is maximum at $l_{ij}\rightarrow l_{0}$ they will preferentially want to accumulate  in regions with lower local stress, as we have also observed in Fig. 3(e) in the main text where the distribution of double bonds around cracks is discussed. This has a direct impact on $P_{u}$, where the second peak at large bond length is reduced. As we have indicated above, the binding probability distribution for immobile and mobile catch bonds are similar.
The unbinding probabilities at higher force are much smaller for catch bonds than for slip bonds, clearly showing that catch bonds are stabilized in high stress regions. 

\newpage

\subsection{Crack identification}

\begin{figure}[!h]
\includegraphics[width=0.8\linewidth]{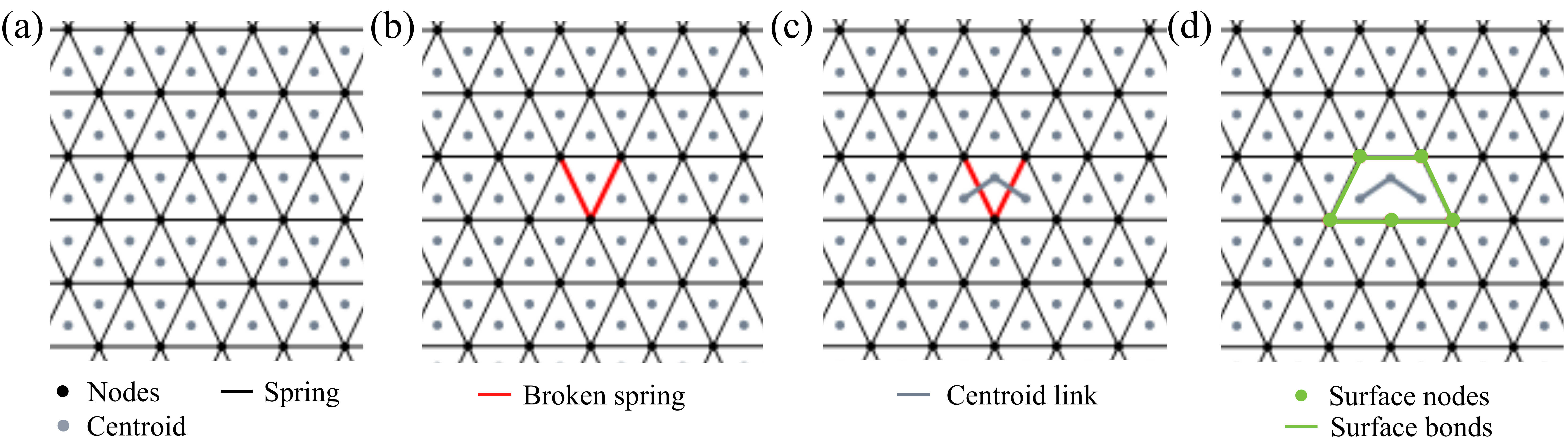}
\caption{Sketch representing the crack identification procedure. (a) Undeformed triangular network, where centroid points of the triangles are represented. (b) Broken bonds are mapped on the undeformed network. (c) Once broken bonds are localized, centroid points are linked, defining a crack. (d) Surface nodes and surface bonds are highlighted around the crack defined on the undeformed network.}
\label{fig:CrackId}
\end{figure}

To identify crack and study crack propagation, we define on the undeformed network, i.e. the fully connected triangular lattice, centroid points of each triangular cell as shown in Fig.~\ref{fig:CrackId}(a). Unbinding and binding events are mapped on the undeformed network, schematically shown in Fig.~\ref{fig:CrackId}(b). Then, two centroid points are linked defining a crack when, during the deformation, the bond that splits them is broken (see Fig.~\ref{fig:CrackId}(c)). Furthermore, we define the crack surface as the lattice nodes or bonds around each crack, as we represent in Fig.~\ref{fig:CrackId}(d). Thus, we compute the crack size distribution $n_{c}$, which we discuss in the main text, as well as the crack surface size distribution $n_{s}$ shown  in Fig.~\ref{fig:SurfaceCrack}(a), exhibiting both the same trend. We also compute the distribution of double bonds around the cracks $P(N_{b,2})$ for different $\sigma$, represented in Fig. 3(e) in the main text.

\subsection{Network evolution}
\subsubsection{Surface size distribution}
\begin{figure}[!h]
\includegraphics[width=\linewidth]{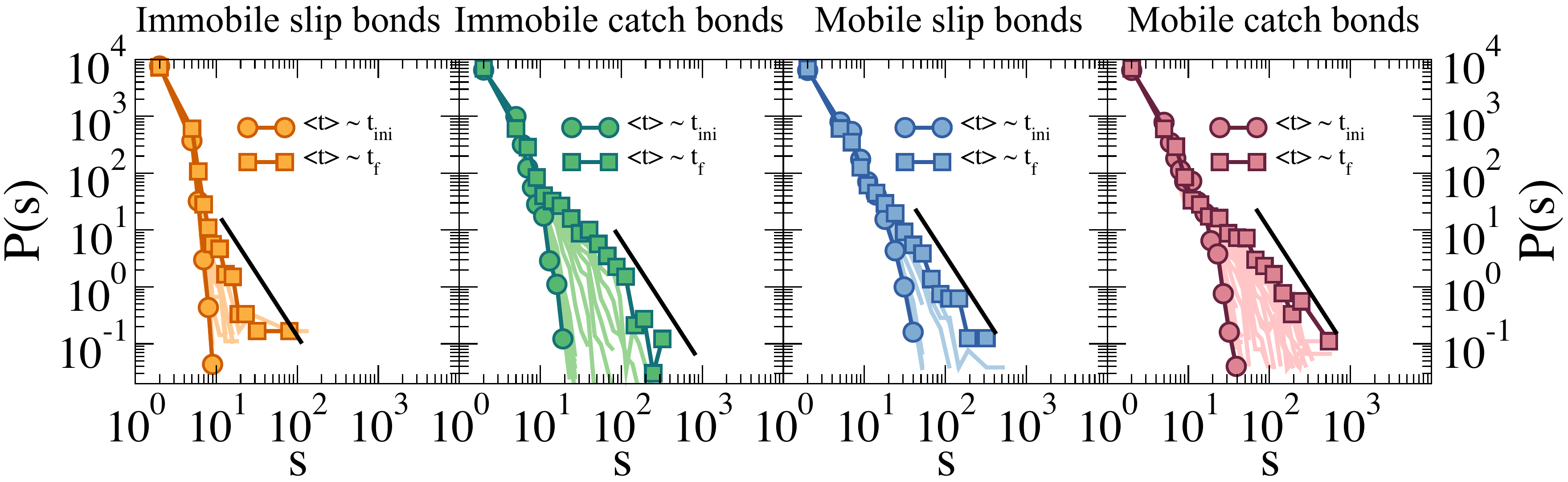}
\caption{Evolution of the surface size distribution $P\left(s\right)$ as a function of the network structure at stress $\sigma=0.36$. In each panel, the full line represents the function $P\left(s\right)\propto s^{-2.2}$. Here, $P\left(s\right)$ is represented at different $t$ in log scale. Only the distributions at early stage ($\left< t\right> \sim t_{ini}$) and close to fracture ($\left< t \right> \sim t_{f}$) are indicated by symbols, to highlight the structural changes.}
\label{fig:SurfaceCrack}
\end{figure}

The crack surface area  distribution $P\left(s\right)$, shown in Fig.~\ref{fig:SurfaceCrack}(a) exhibits the same behaviour as the the crack size distribution represented in Fig. 3 in the main text. Indeed, $P\left(s\right)$ shows a power-law dependence at large crack sizes with an exponent $\alpha\approx-2.2$ just before the network fractures for networks with immobile slip, immobile catch and mobile catch bonds. This value is consistent with the random percolation value~\cite{torquato2002random,sciortino2005one}. In addition, both $P\left(c\right)$ and $P\left(s\right)$ evolve in the same way as a function of $t$. 

\subsubsection{Network evolution at low stress}

\begin{figure}[!h]
\includegraphics[width=0.8\linewidth]{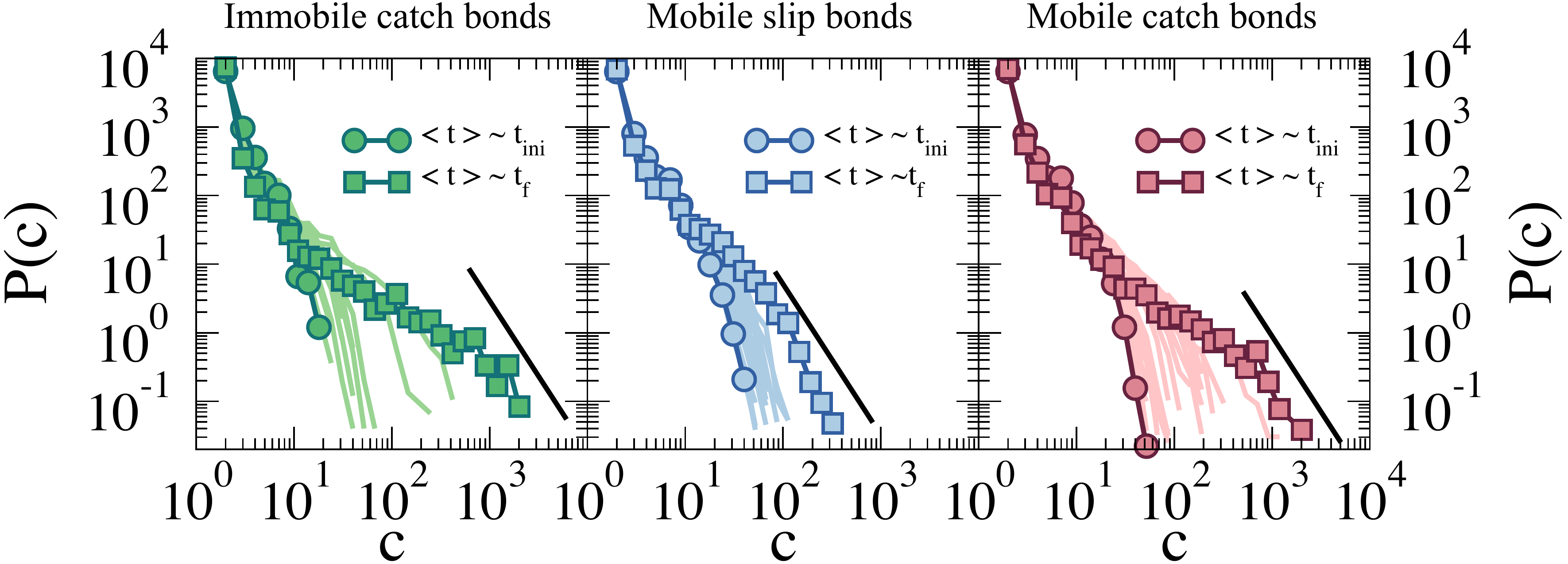}
\caption{Evolution of the crack size distribution $P\left(c\right)$ as a function of the network structure at stress $\sigma=0.10$, for immobile catch bonds, mobile slip bonds, and mobile catch bonds. In each panel, the full line represents the function $P\left(c\right)\propto c^{-2.2}$. Here, $P\left(c\right)$ is represented at different $t$ in log scale, although only at early stage ($\left< t\right> \sim t_{ini}$) and close to fracture ($\left< t \right> \sim t_{f}$) are indicated by symbols to highlight the structural changes.}
\label{fig:Cluster010}
\end{figure}

In Fig.~\ref{fig:Cluster010} we show $P_{c}$ at $\sigma=0.10$ for networks with immobile slip bonds, immobile catch bonds and mobile catch bonds (note that networks with immobile slip bonds do not break at this stress). The distributions look similar to those at higher stress in Fig. 3 in the main text, albeit it that the distributions for catch bonds are even wider at this lower stress.

\subsubsection{Double bond distribution}

\begin{figure}[!h]
\includegraphics[width=0.8\linewidth]{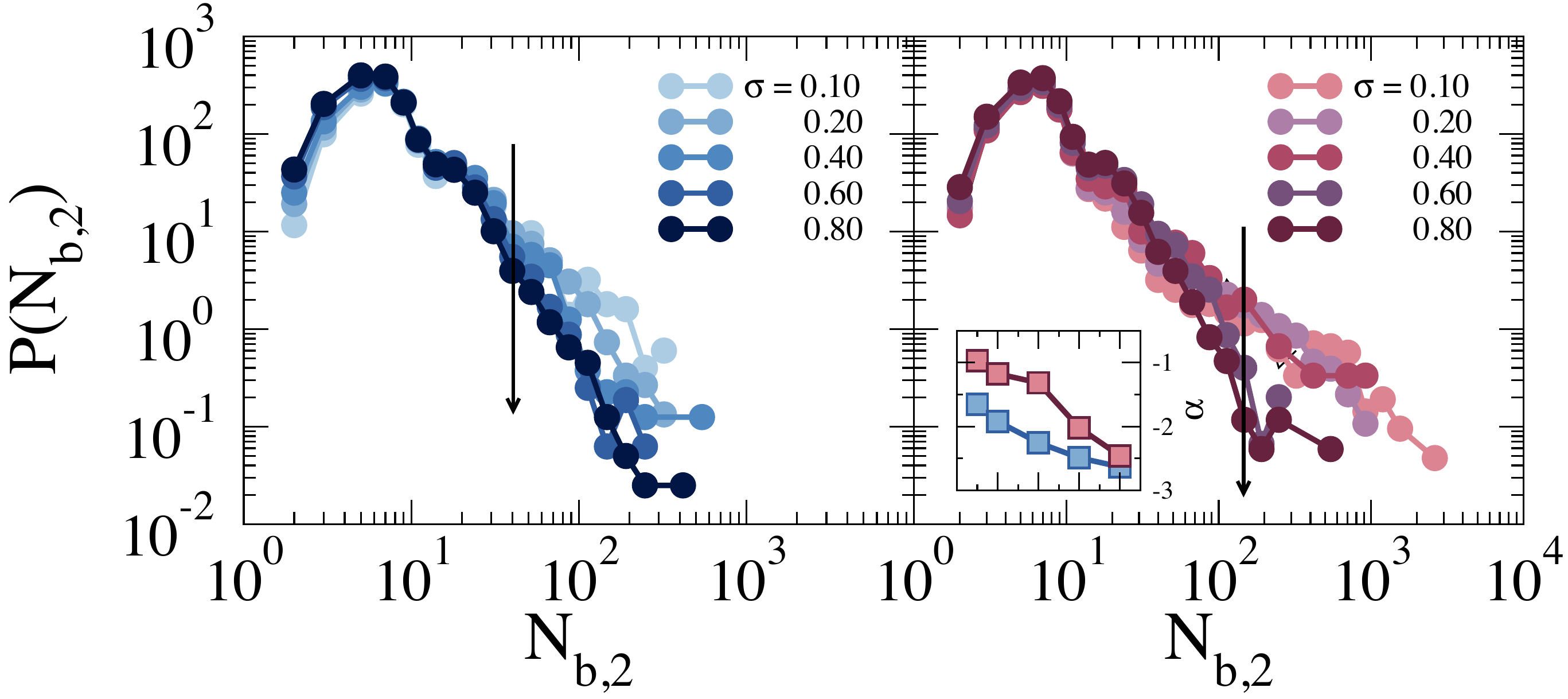}
\caption{Distribution of double bonds $P\left(N_{b,2}\right)$ computed on the crack surface as a function of $\sigma$ close to the moment of fracture $t_{f}$, for networks with mobile slip bonds (blue) and mobile catch bonds (red). The right arrow highlights the stress direction. {\it Inset:} Exponent corresponding to the fitting $P\left(N_{b,2}\right)\propto N_{b,2}^{-\alpha}$ made to the tail of the distribution.}
\label{fig:DoubleBondsStress}
\end{figure}

In Fig.~\ref{fig:DoubleBondsStress} we show the distribution of double bonds $P\left(N_{b,2}\right)$ computed on the crack surface for different $\sigma$ for networks with mobile slip and mobile catch bonds. In particular, for mobile slip bonds we can see that the distribution decays faster by increasing $\sigma$, revealing that bonds are less stable due to the high local stress on the crack surface. By contrast, the effect of stress on $P\left(N_{b,2}\right)$ is smaller for networks with catch bonds. We fit the tail of these distributions by considering $P\left(N_{b,2}\right)\propto N_{b,2}^{-\alpha}$. Te inset of Fig.~\ref{fig:DoubleBondsStress} clearly shows that the accumulation of catch bonds is more pronounced than for slip bonds. Thus, crack propagation is hindered more efficiently, postponing the fracture because there is an accumulation of bonds around the cracks.

\newpage

\subsubsection{Damage visualization}

\begin{figure*}[!h]
\includegraphics[width=0.8\linewidth]{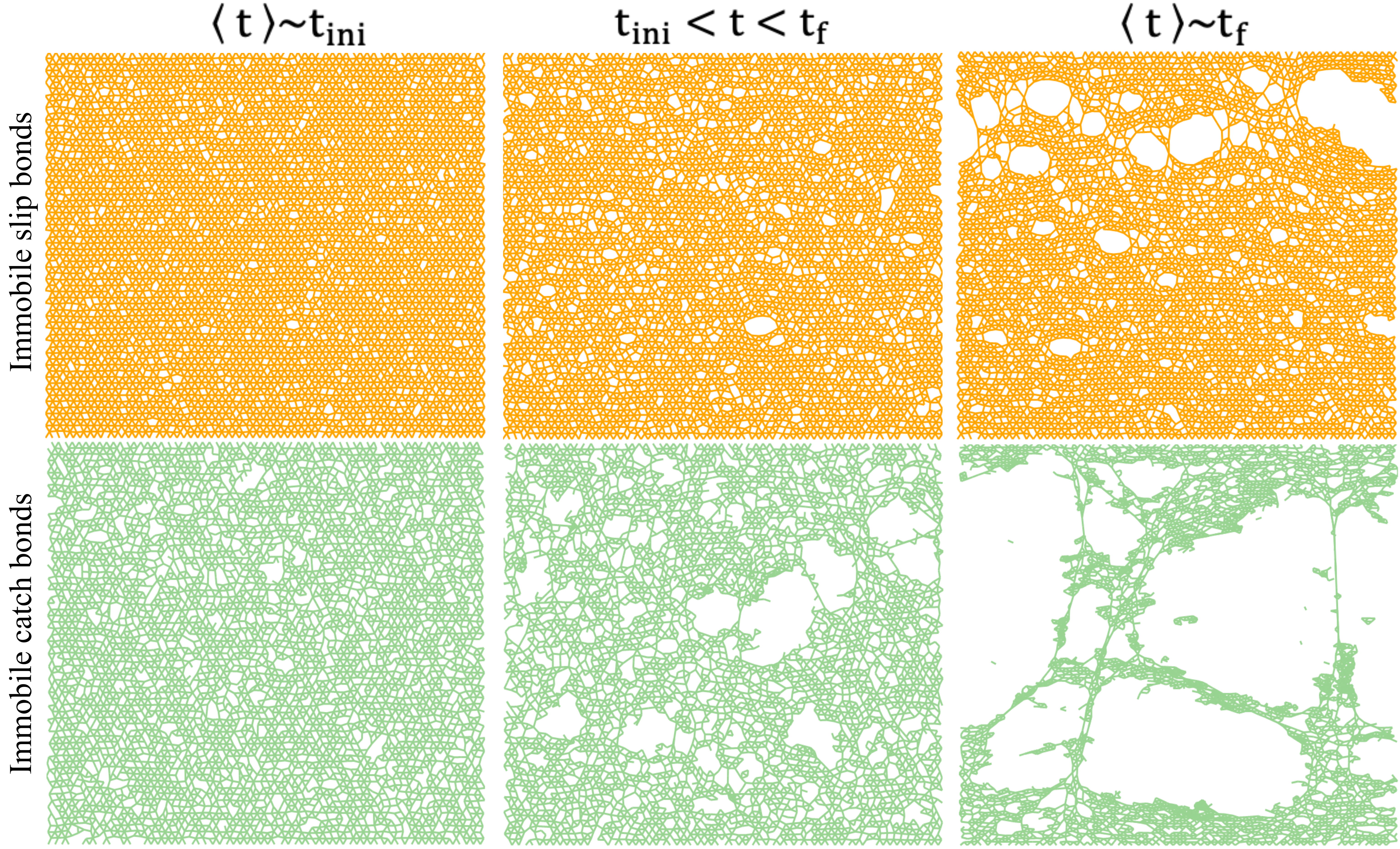}
\caption{Network evolution as a function of time $t$ with immobile bonds, for $\sigma=0.40$.}
\label{fig:Evol_imm}
\end{figure*}

\begin{figure*}[!h]
\includegraphics[width=0.8\linewidth]{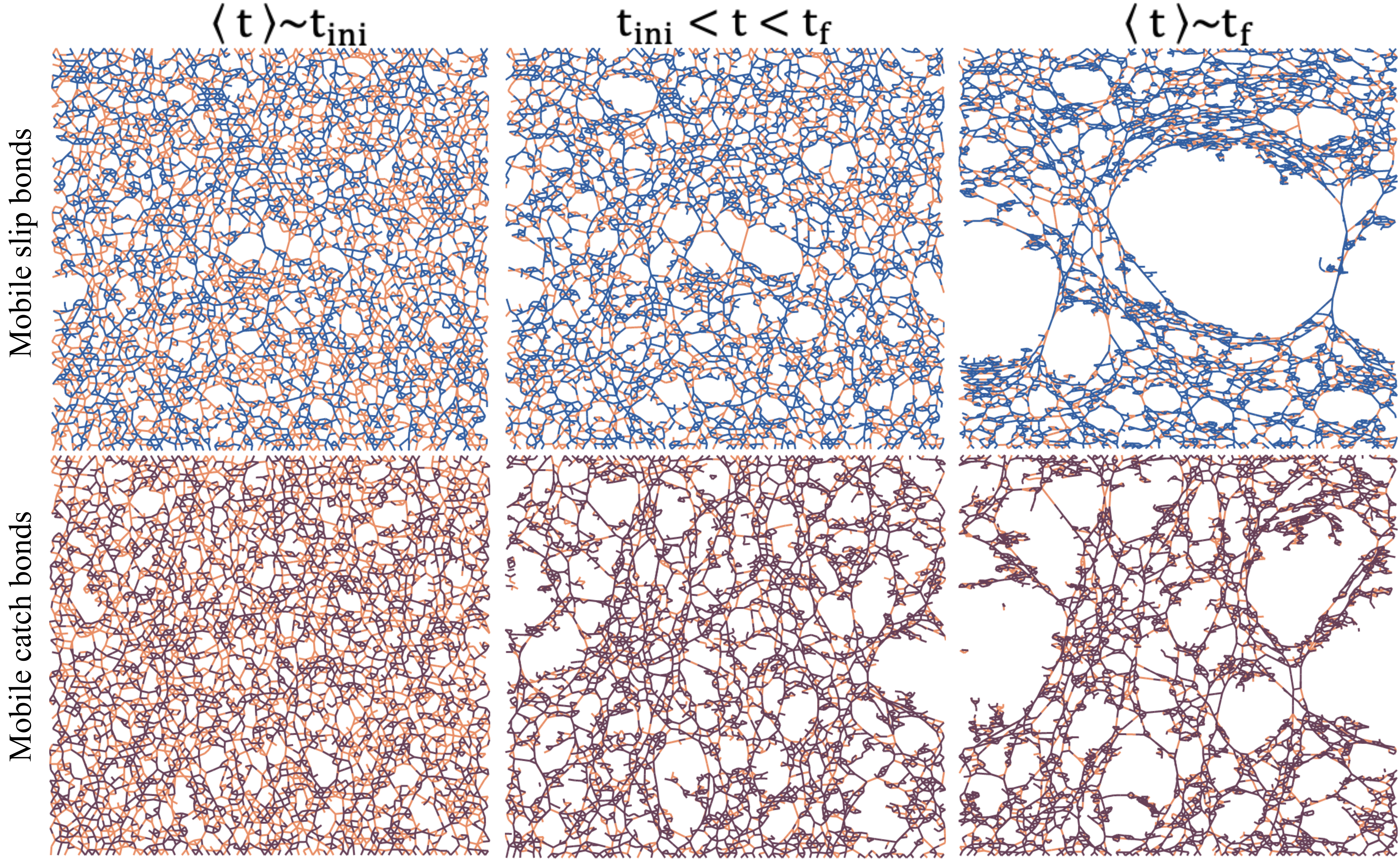}
\caption{Network evolution as a function of time $t$ with mobile bonds, for $\sigma=0.40$. Orange indicates single bonds, whereas blue and purple indicate double slip and catch bonds, respectively.}
\label{fig:Evol_mob}
\end{figure*}

The evolution discussed by computing $P\left(c\right)$ at $\sigma=0.40$ is represented by snapshots taken at different times for networks with immobile bonds in Fig.~\ref{fig:Evol_imm} and for networks with mobile bonds in Fig.~\ref{fig:Evol_mob}, where single bonds are shown in orange, and double slip and catch bonds are shown in blue and purple, respectively. Likewise, we also visualize the damage propagation on the network once the fracture has taken place, i.e. $t=t_{f}$.

\begin{figure*}[!h]
\includegraphics[width=\linewidth]{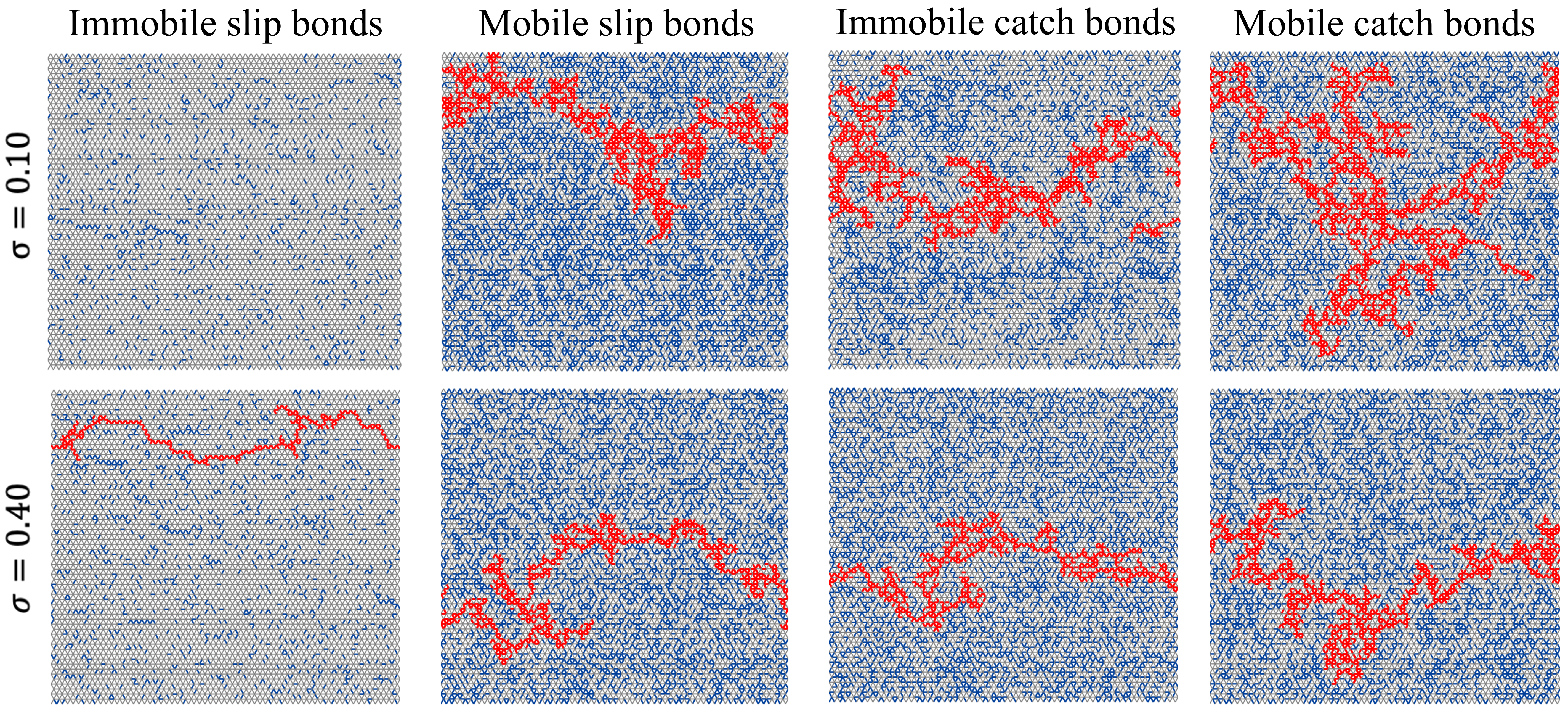}
\caption{Visualization of the damage once the fracture has taken place, i.e. $t=t_{f}$, mapped on the undeformed network at $\sigma=0.10$ (top) and $\sigma=0.40$ (bottom). Here, broken bonds belonging to the percolated crack are highlighted in red color, whereas blue color indicates individual broken bonds. Note that the network with immobile slip bonds does not fracture for $\sigma=0.10$.}
\label{fig:Evol_imm2}
\end{figure*}

\clearpage
\bibliographystyle{unsrt}
\bibliography{apssamp}

\end{document}